\documentclass[]{spie}  
\usepackage[dvips]{graphicx}

\newcommand{\NH}{\mbox {$N_{\rm H}$}}
\newcommand{\etal}{\mbox{{\em et~al.}}}

\title{The Flight Spectral Response of the ACIS Instrument}

\author{Paul P. Plucinsky\supit{a}, Norbert S. Schulz\supit{b}, 
Herman L. Marshall\supit{b}, Catherine E. Grant\supit{b}, \\
George Chartas\supit{c}, Divas Sanwal\supit{c}, Marcus
A. Teter\supit{c}, Alexey A. Vikhlinin\supit{a}, \\
Richard J. Edgar\supit{a}, Michael W. Wise\supit{b}, Glenn
E. Allen\supit{b}, Shanil N. Virani\supit{a}, \\
Joseph M. DePasquale\supit{a}, and 
Michael T. Raley\supit{a} 
\skiplinehalf
\supit{a}Harvard-Smithsonian Center for Astrophysics, 60 Garden St., 
Cambridge, MA 02138 \\ 
\supit{b}Center for Space Research, Massachusetts Institute of
Technology,  Cambridge, MA 02139 \\
\supit{c}Department of Astronomy \& Astrophysics, Pennsylvania State 
University, \\
University Park, PA 16802 }

\authorinfo{Further author information: (Send correspondence to
P.P.P.)\\P.P.P.: E-mail: plucinsky@cfa.harvard.edu, Telephone: 1 617
496 7726\\  }

\begin{document} 
  \maketitle 

\begin{abstract}
We discuss the flight calibration of the spectral response of 
the {\em Advanced CCD Imaging Spectrometer} (ACIS) 
on-board the {\em Chandra X-ray Observatory} (CXO).  The spectral
resolution and sensitivity of the ACIS instrument have both
been evolving over the course of the mission. The spectral
resolution of the frontside-illuminated (FI) CCDs changed
dramatically in the first month of the mission due to
radiation damage.  Since that time, the spectral resolution
of the FI CCDs and the backside-illuminated (BI) CCDs
have evolved gradually with time.  We demonstrate the efficacy of
charge-transfer inefficiency (CTI) correction algorithms which
recover some of the lost performance.  The detection efficiency of
the ACIS instrument has been declining throughout the
mission, presumably due to a layer of contamination 
building up on the filter and/or CCDs.  We present a
characterization of the energy dependence of the excess 
absorption and demonstrate software which models the time
dependence of the absorption from energies of 0.4~keV and up.   
The spectral
redistribution function and the detection efficiency are well-characterized
at energies from 1.5 to 8.0~keV primarily due to the existence
of strong lines in the ACIS calibration source in that
energy range.  The calibration at energies below 1.5 keV is
challenging because of the lack of strong lines in the
calibration source and also because of the inherent non-linear
dependence with energy of the CTI and the absorption by the
contamination layer.
We have been using data from celestial sources with relatively
simple spectra to determine the quality of the calibration
below 1.5 keV.  We have used observations of 1E0102.2-7219 
(the brightest supernova remnant in the SMC), PKS2155-304
(a bright blazar), and the pulsar PSR~0656+14 (nearby pulsar with a
soft spectrum), since the spectra of these objects have been
well-characterized by the gratings on the CXO.  The analysis 
of these observations demonstrate that the CTI correction
recovers a significant fraction of the spectral resolution of the
FI CCDs and the models of the time-dependent absorption result in 
consistent measurements of the flux at low energies for data from a
BI~(S3) CCD.


\end{abstract}



\section{INTRODUCTION}
\label{sect:intro}  

The Chandra X-ray Observatory (CXO) is the third of NASA's great
observatories in space\cite{weiss00,weiss02}.  The CXO was launched just
past midnight on 
July 23, 1999 aboard the space shuttle \textit{Columbia} on the STS-93
mission.  The CXO was placed into a higher orbit by an Inertial Upper
Stage (IUS) booster and then used its own propulsion system to achieve
a final orbit with a perigee of 10,000~km, an apogee of 140,000~km,
an inclination of $28.5^\circ$ and a period of $\sim64$~hr.
The CXO is controlled and operated by the Smithsonian Astrophysical 
Observatory (SAO) from Cambridge, Massachusetts.  The Chandra X-ray
Center (CXC), also run by the
Smithsonian Astrophysical Observatory, processes and distributes
Chandra data and provides analysis
software and calibration products to the astronomical community.

The CXO carries two focal plane science instruments: the \textit{Advanced 
CCD Imaging Spectrometer} (ACIS) and the \textit{High Resolution
Camera} (HRC). 
The Observatory also possesses two objective transmission
gratings: a \textit{Low Energy Transmission Grating} (LETG) that is  
primarily used with the HRC, and the \textit{High Energy Transmission Grating}
(HETG) that is primarily used with the ACIS. 
ACIS was developed by a team from the Massachusetts Institute of
Technology\cite{bautz98} and the Pennsylvania State 
University \cite{garmire92} and 
is the primary scientific instrument aboard CXO, currently 
conducting $\sim90\%$ of the observations. It contains two arrays of
CCDs, one optimized for imaging and the other for spectroscopy as the
readout detector for the HETG.  The ACIS imaging array contains 4 
Front-Illuminated (FI) CCDs configured in a $2\times2$ array and the
spectroscopy array contains 2 Back-Illuminated (BI) CCDs and 4~FI~CCDs
configured in a $1\times6$ array.  

\section{THE TIME EVOLUTION OF THE SPECTRAL RESPONSE}

The spectral response of the ACIS CCDs has been evolving with 
time since the launch of CXO due to an increase in the 
charge-transfer inefficiency (CTI) and the development of a
contamination layer on the filter and/or the
CCDs.     The CTI introduces a spatial dependence in the spectral 
resolution of the detectors. The contamination layer introduces an 
energy-dependent reduction in the detection efficiency of the instrument.
The relative importance of these effects
are different for the FI and BI CCDs.  We discuss these effects
in detail in the following sections and also discuss the
the low-energy gain and spectral redistribution function of the 
BI CCDs.

   \begin{figure}[h]
   \vspace{-0.50in}
   \begin{center}
   \begin{tabular}{c}
   \includegraphics[width=4in,angle=270]{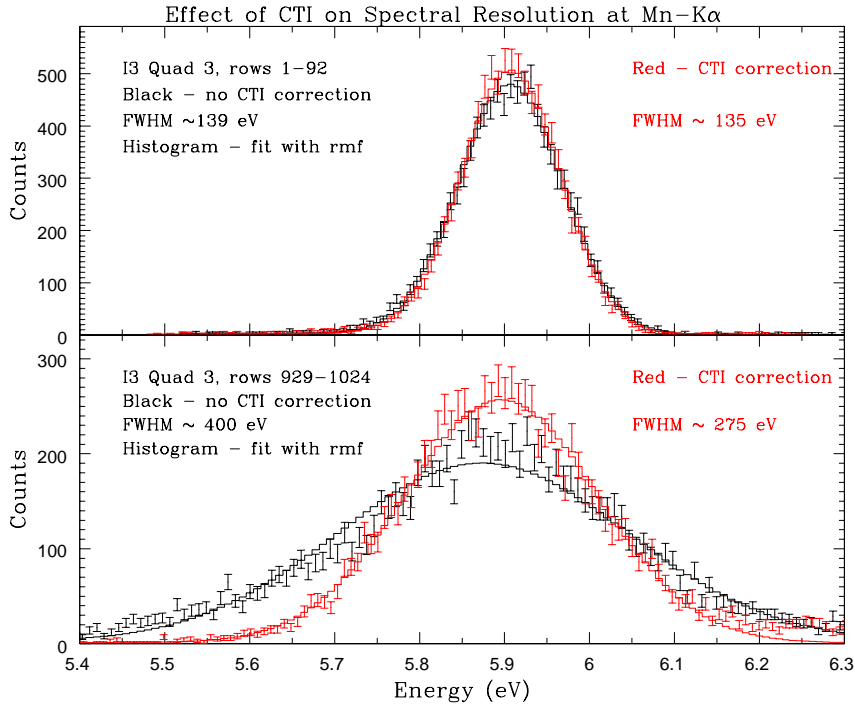}
   \end{tabular}
   \end{center}
   \vspace{-0.15in}
   \caption[Mn_PHdist] 
   { \label{fig:Mn_PHdist}
Comparison of PH distributions from Mn ${\rm K}\alpha$ X-rays
(5.9~keV) from the
ACIS external calibration source with and without a CTI correction. The
top panel displays the PH
distribution from the bottom of the CCD where the effects of CTI are
minimal. The bottom panel displays the PH distribution from the top of
the CCD where the effects of CTI are at a maximum.
}
   \end{figure} 

\subsection{RADIATION DAMAGE AND CHARGE TRANSFER INEFFICIENCY}

In the first month of the mission, the FI CCDs on ACIS suffered
a large increase in CTI\cite{prig00a,prig00b} (CTI is defined as the
fractional charge loss per pixel transferred).  The CTI increased from 
essentially zero to $\sim1.0-2.0\times10^{-4}$.  The CTI varies
from CCD to CCD as the FI CCDs in the imaging array were less
damaged than the FI CCDs in the spectroscopy array.  The BI CCDs
showed no measurable increase in CTI during this time frame.
During the first month of the mission, ACIS was left at the focus
of the telescope during the radiation belt transits.  It is believed
that the damage was the result of low-energy protons ($\sim 100-200$
keV) which scattered off of the CXO's mirrors with a sufficient 
efficiency to produce a significant flux at the focal
plane.  The ACIS instrument is now moved out of the focal position of
the telescope during every radiation belt transit and the rapid
increase in CTI has not continued.  The CTI of both the FI and BI CCDs
has been slowly increasing over the remaining three years of the
mission.  This gradual increase is presumably the result of the
accumulated damage of high-energy protons which penetrate the ACIS 
proton shield and residual low-energy protons which are encountered
outside of the radiation belts.

  The ACIS CCDs are framestore design CCDs. The CCD consists of
an imaging array and a framestore array.  The imaging array is 
exposed to the incident radiation and the framestore array is 
covered by an Al shield. The framestore cover is sufficiently thick
to stop the low-energy protons which produce the increase in CTI.
Therefore, the CTI of the framestore array in the FI CCDs is unchanged since
launch and is still essentially zero.  Figure~\ref{fig:Mn_PHdist}
displays the effect of CTI on a pulse-height (PH) distribution 
generated from Mn ${\rm K}\alpha$ X-rays from the ACIS external
calibration source in flight on the I3(FI) CCD.  The top panel shows
the PH distribution
from rows 1-92 of the CCD near the framestore array and the bottom
panel shows the PH distribution from rows 929-1024 near the top of the
CCD. Since only the parallel CTI in the imaging array increased and
the parallel and serial CTI in the framestore array remained
unchanged, the problem was reduced to one dimension
in row number on the CCD.  It was possible to construct an algorithm
which compensated for the charge lost to CTI on an event-by-event
basis which improves the resulting spectral resolution~\cite{town00,town02a}.
The correction software recovers some, but not all, of the lost
performance.  In Figure~\ref{fig:Mn_PHdist}, we have plotted the PH
distributions after the CTI correction software (SW) has been applied to the 
data. Note that the correction modifies the data only slightly near
the framestore region, but modifies the data significantly near the
top of the CCD.  The correction algorithm improves the spectral
resolution  by $\sim30$\% at the top of the CCD.

   \begin{figure}[h]
   \vspace{-0.30in}
   \begin{center}
   \begin{tabular}{c}
   \includegraphics[width=2.5in,angle=270]{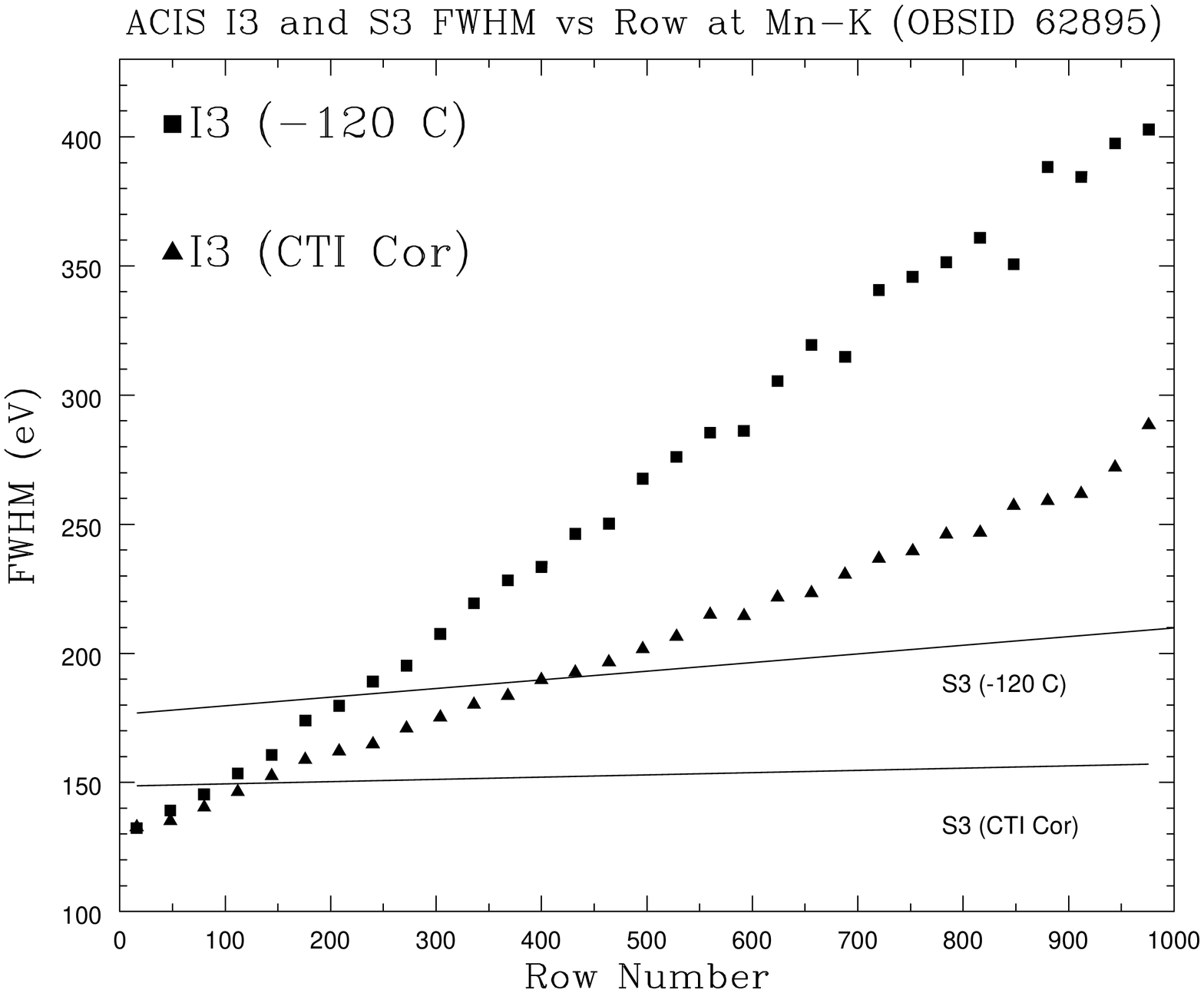}
   \includegraphics[width=2.5in,angle=270]{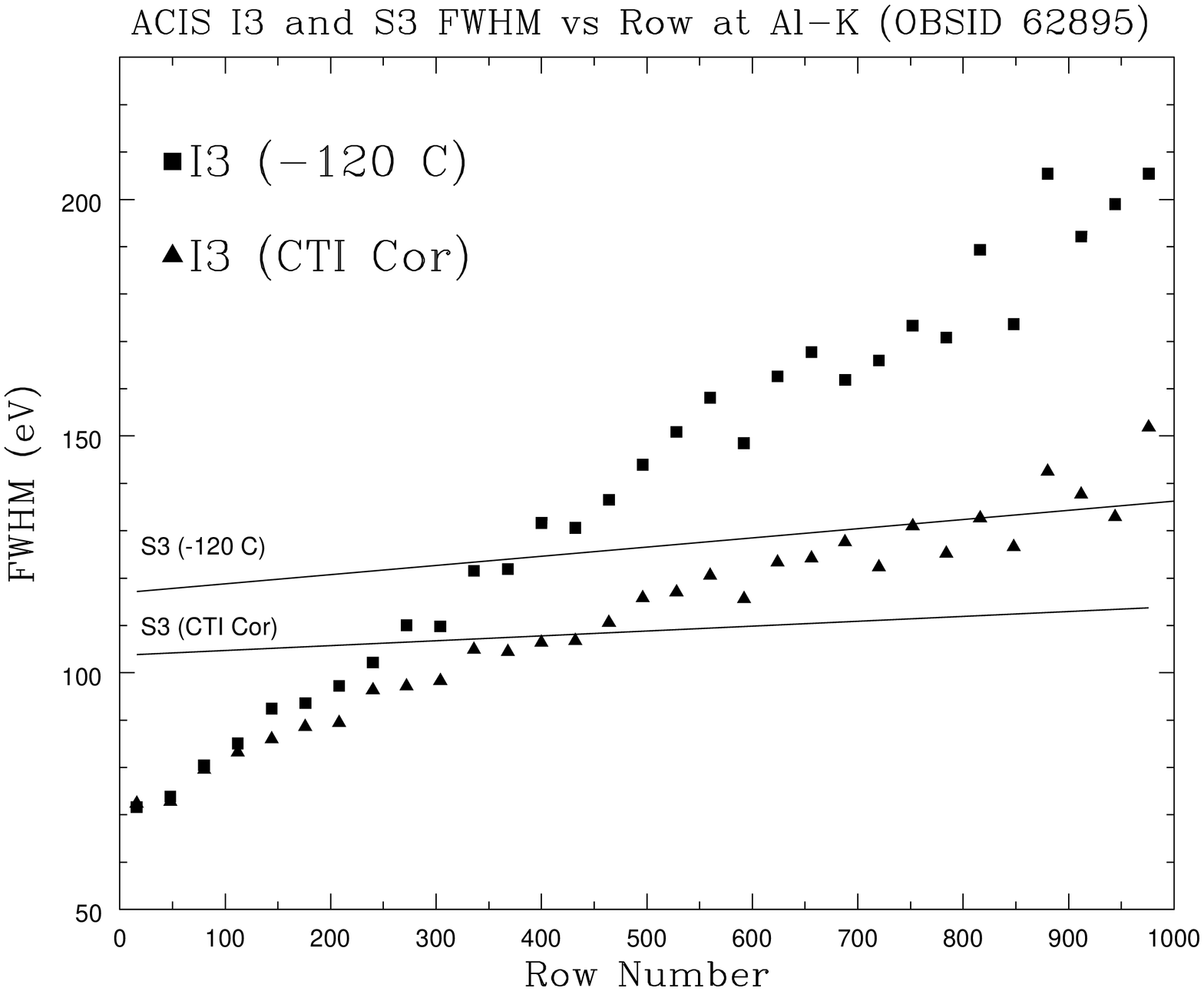}
   \end{tabular}
   \end{center}
   \vspace{-0.15in}
   \caption[Mn_FWHM] 
   { \label{fig:Mn_FWHM}
The FWHM of the ACIS I3 detector versus row number with and without
a CTI correction for Mn ${\rm K}\alpha$ X-rays (5.9~kev) (LEFT) and 
Al ${\rm K}\alpha$ X-rays (1.5~keV) (RIGHT).  Complementary 
numbers for the S3 detector are plotted for reference.
}
   \end{figure} 

We have analyzed the data from a long observation under the ACIS
external calibration source (OBSID 62895) to determine the variation
of the full-width at half maximum (FWHM) versus row number.  We have 
conducted this analysis with the
uncorrected data and the CTI-corrected data.  Figure~\ref{fig:Mn_FWHM}
displays the FWHM versus row number for Mn ${\rm K}\alpha$ X-rays and
for Al ${\rm K}\alpha$ X-rays for the I3(FI) CCD.  The CTI
correction provides a significant improvement at both energies.
We have also plotted the FWHM versus row number for the S3 CCD
for comparison.  After applying the CTI correction, the FI CCDs
in the ACIS imaging array
provide better spectral resolution than
S3 over the bottom half of the CCD at 1.5~keV.

 We have used an implementation of the CTI corrector
developed by the CXC for a future release of the {\em Ciao}
SW for the I3 data and have used the publicly available CTI correction
SW from the Penn State group for the S3 data.  We have compared the I3
results from the CXC CTI corrector and the Penn State CTI corrector
and found the results to be nearly identical.  The CTI correction SW
is available from the
Penn State group at the CXC contributed SW web site:\\
{\tt
http://asc.harvard.edu/cont-soft/software/ACISCtiCorrector.1.37.html}.


\subsection{CONTAMINATION BUILDUP AND LOW-ENERGY DETECTION EFFICIENCY}
\label{sect:contam}

\begin{figure}[h]
   \vspace{-0.30in}
   \begin{center}
   \begin{tabular}{c}
   \includegraphics[width=4.0in,angle=270]{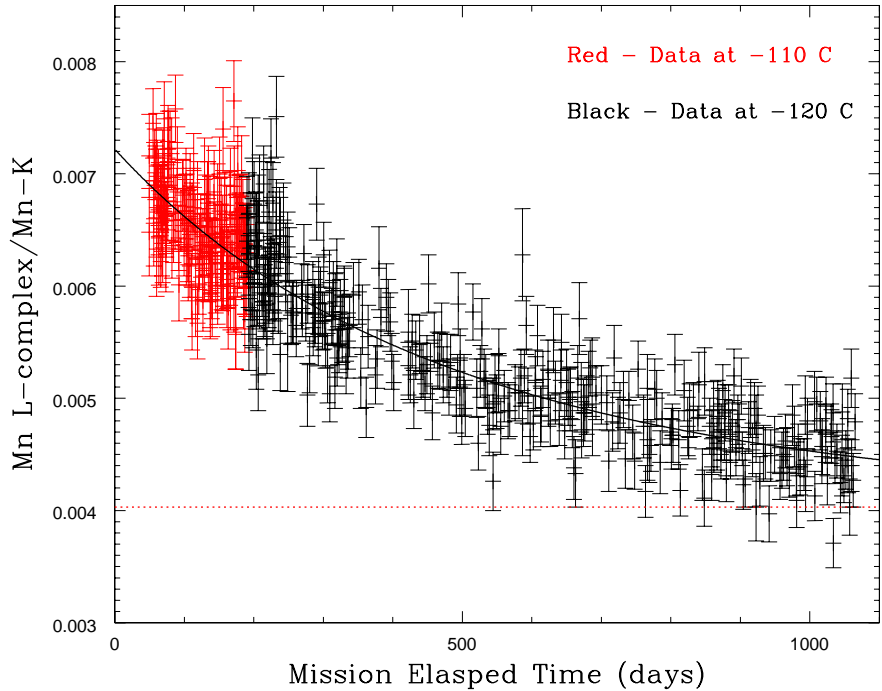}
   \end{tabular}
   \end{center}
   \vspace{-0.15in}
   \caption[MnLK_ratio] 
   { \label{fig:MnLK_ratio}
The ratio of the Mn L/Fe L complex/Mn K lines versus time from the
S3(BI) CCD.  The data collected
with the focal plane at -110 C are indicated in red
and the data collected at -120 C are indicated in black.
The solid line is a model of the time dependence in 
Eqn.~\ref{eq:ratio} and the red dashed line is the asymptotic value of
0.00403.
}
   \end{figure}

  It was noticed in early 2002 that the detection efficiency of the ACIS
instrument had been declining gradually over the life of the mission.
The effect at energies above 1.5~keV is negligible, but the effect at
lower energies is large ($\sim 50$\% reduction in quantum efficiency
at 0.5 keV over the first three years of the mission).  The ACIS
external calibration source produces weak
Mn~L and Fe~L lines in addition to the Mn~K lines.  We have analyzed the ratio
of the Mn~L complex and Mn~K lines over the life of the mission.  The data are
plotted in Figure~\ref{fig:MnLK_ratio} and indicate the ratio has
been decreasing throughout the mission.  The rate of decrease appears to
be slowing with time.  The data have been fit with a function of the 
form\cite{odell02}:

\begin{equation}
        \label{eq:ratio}
{\rm Mn~L/Mn~K~ratio} = {\rm N_o}\times exp[ -\tau_{\infty}\times(1-exp(-t/\tau_o))]
\end{equation}

\noindent where ${\rm N_o}$ is the value of the ratio at $t=0$, $t$ is the
number of days since launch, $\tau_{\infty}$ is the value of the
optical depth at $t=\infty$, and $\tau_o$ is a characteristic time
constant.  Since this function is an exponential of an exponential
decay in time, we cannot interpret the value of $\tau_o$ as the
characteristic time constant of the overall decay.
The asymptotic value of this function is 0.00403.  If this model can
be legitimately extrapolated into the future, this implies that 
$\sim80$\% of the decrease in the ratio has already occurred.

%
   \begin{figure}[h]
   \vspace{-0.30in}
   \begin{center}
   \begin{tabular}{c}
   \includegraphics[width=5in,angle=0]{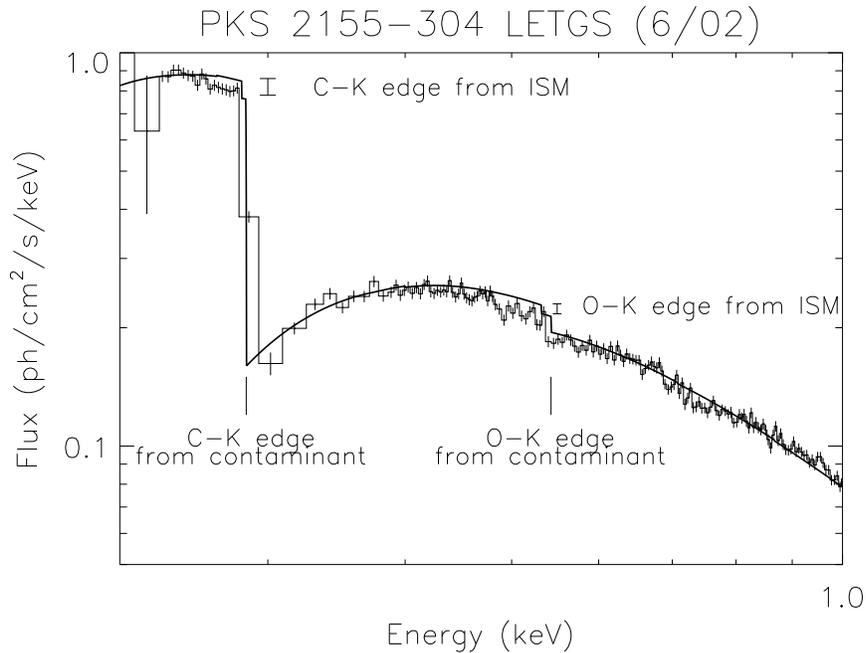}
   \end{tabular}
   \end{center}
   \vspace{-0.15in}
   \caption[PKS2155_LETG] 
   { \label{fig:PKS2155_LETG}
The ACIS LETG spectrum of PKS2155 from OBSID 3669. The C and O
absorption edges from the ISM and the contaminant are identified.
}
   \end{figure} 

   \begin{figure}[h]
   \vspace{-0.30in}
   \begin{center}
   \begin{tabular}{c}
   \includegraphics[width=4.3in,angle=270]{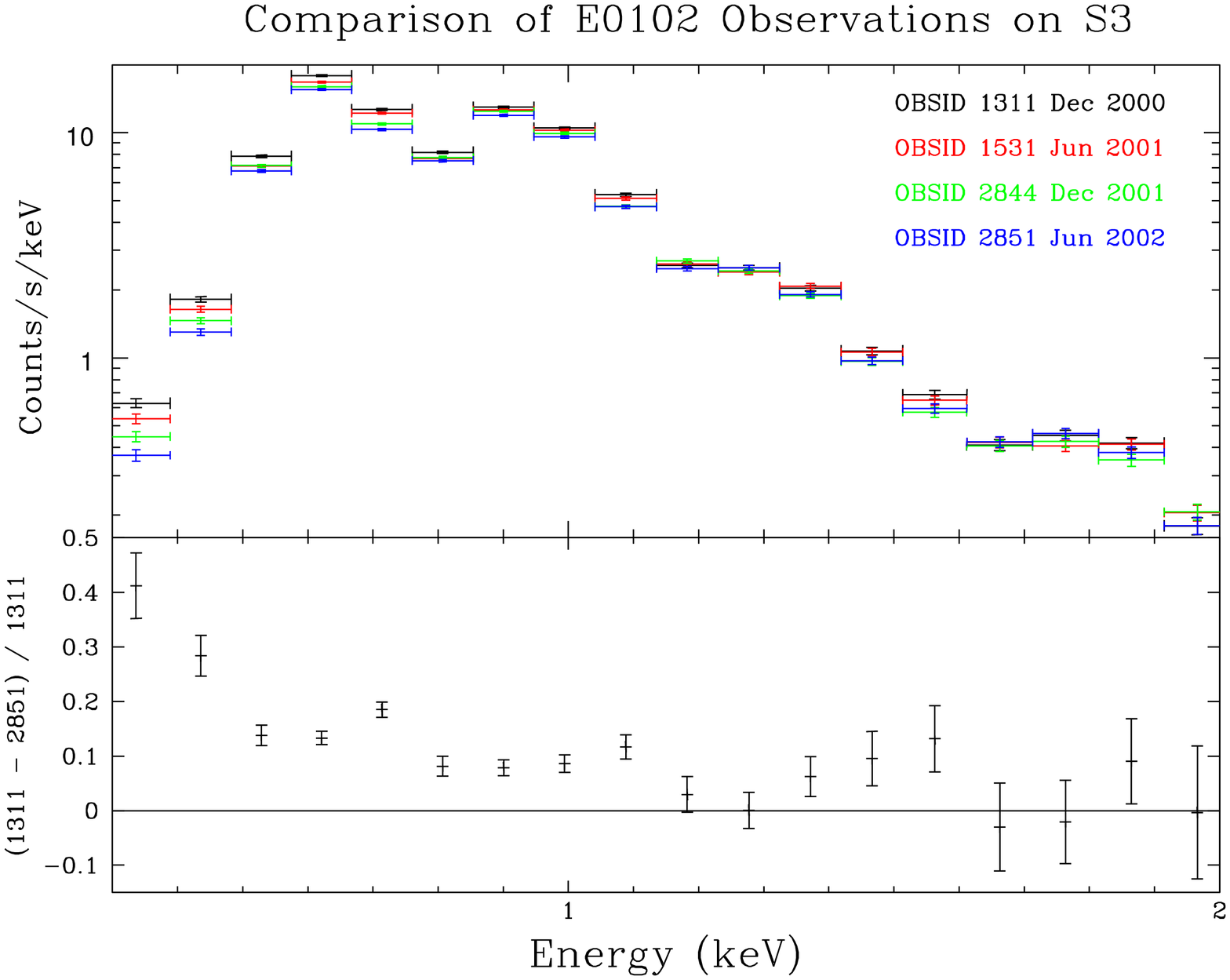}
   \end{tabular}
   \end{center}
   \vspace{-0.50in}
   \caption[E0102_time] 
   { \label{fig:E0102_time}
S3(BI) observations of E0102 at different epochs, different epochs are
indicated by different colors.  The data have been
binned into relatively large bins in energy space to provide a
statistically significant number of counts in each bin.  The bottom
panel shows the fractional difference between the earliest and latest
observations.
}
   \end{figure} 

 Once the putative contamination layer was discovered in early 2002, a
series of calibration observations were conducted to better
characterize the nature of the contamination.  Several observations
were conducted in the LETG/ACIS configuration to provide a
well-characterized spectrum with a large flux of low energy photons.
Figure~\ref{fig:PKS2155_LETG} displays the LETG/ACIS spectrum from
OBSID~3669, an 45ks observation of the bright blazar PKS2155-304.

The spectrum of PKS 2155-304 was fit with two power-law components
absorbed by neutral gas in the Galactic ISM using the abundances
and opacities of Wilms~\etal\cite{wilms00}.  The C-K and O-K edges due to the
contaminant were fit to the data, giving optical depths of
$1.58\pm0.02$ and $0.10\pm0.01$, respectively.  Residuals of $\sim5\%$
near 0.5~keV probably result from a poor model of the transmission of
the contaminant.  There is no detection of a N-K edge and the 90\%
upper limit to the optical depth there is about 0.04.  Residuals near
0.7~keV may result from fluorine in the contaminant but there is
no clear edge nor is there a good model of the absorption due to
fluorine in relevant compounds.

Since the contamination layer is composed mostly of C, the largest
effect on CXO data will be at energies below 1.0~keV.  We have
examined calibration observations of 1E0102.2-7219, the brightest
supernova remnant in the Small Magellanic Cloud which has a soft,
line-dominated spectrum.  Figure~\ref{fig:E0102_time} displays four
observations of 1E0102.2-7219 separated by two years in time.
The bottom panel of the plot displays the difference between the 
earliest and latest observation.  It is clear from these data that
the largest effects occur at energies below 1.5~keV.  It is also
clear that the data above 1.5~keV are mostly unaffected.

\subsection{LOW-ENERGY GAIN AND SPECTRAL REDISTRIBUTION FUNCTION FOR
THE BI CCDS }

 We have reported  previously on our efforts to improve
the response model of the BI CCDs on ACIS~\cite{ppp01}.  For a 
thorough discussion of the models developed for the ACIS CCDs, the
reader is referred to the publications by the ACIS 
team\cite{bautz99,town02b}.  
The model of the spectral redistribution function for the 
S3 response matrix was significantly improved in the release of the
CXO calibration database CALDB2.7.  At that time we noted that the
gain and the redistribution function were well-modeled above 0.8 keV,
but that there were still deficiencies below 0.8 keV.
  We have used the observations of PKS2155-304 with the LETG to
investigate the response of the S3 CCD.  The advantage of the LETG
data is that the energy of the incident photons can be determined from
their position on the CCD and can therefore be used to verify the
conversion from pulse height to energy.
By using pointing offsets in multiple
observations, we have been able to acquire a range of energies at
the same location on the CCD.  These data can then be used to
characterize the gain and spectral redistribution function at low
energies.

   \begin{figure}[h]
   \vspace{-0.30in}
   \begin{center}
   \begin{tabular}{c}
   \includegraphics[width=4.5in,angle=0]{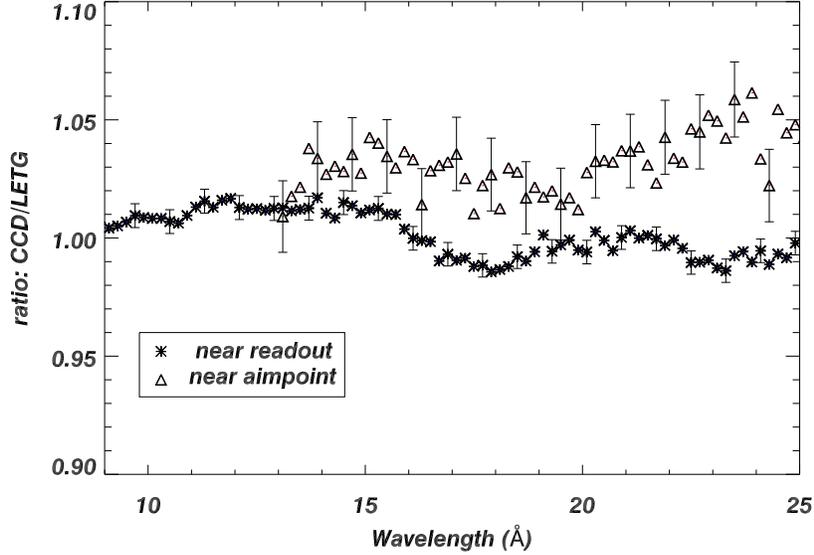}
   \end{tabular}
   \end{center}
   \vspace{-0.15in}
   \caption[nss_gain] 
   { \label{fig:nssgain}
Evaluation of the low-energy gain of the S3 CCD.  The ratio of the
energy determined by the PH of the event and the energy determined by
the dispersion of the LETG is plotted versus wavelength.  Two
locations on the CCD are plotted, one near the aimpoint (OBSIDs 2323
\& 2324) and one near the framestore (OBSID 3669) in chip row number.
}
   \end{figure} 

  We have used the data from OBSIDs 2323, 2324, and 3669 to
investigate the gain at low energies.  OBSIDs 2323 and 2324 positioned
the dispersed spectrum from the LETG at the nominal readout location
in the middle of the CCD and OBSID 3669 positioned the dispersed
spectrum closer to the framestore region using a translation offset
in the Science Instrument Module (SIM).  Figure~\ref{fig:nssgain}
plots the ratio of the energy determined from the PH of the event
to the energy determined by the dispersion relation of the LETG.
The energy determined from the PH of the event agrees with the
LETG-determined energy to within $\pm2$\% near the readout. However,
the data near the aimpoint show a systematic offset of $2-6$\% for
energies below 900~eV with 
the CCD PH overestimating the true energy of the photon.  We will
be able to use the multiple calibration observations of PKS2155-304
to refine the low-energy gain of the S3 CCD as a function of position
to be incorporated in an updated matrix of the CXO CALDB.  It is
important to emphasize the energy of the events as determined by the 
PH agrees with the LETG-determined energy at both locations for
energies above 0.9~keV.

\begin{figure}[h]
   \begin{center}
   \begin{tabular}{c}
   \includegraphics[width=2.7in,angle=270]{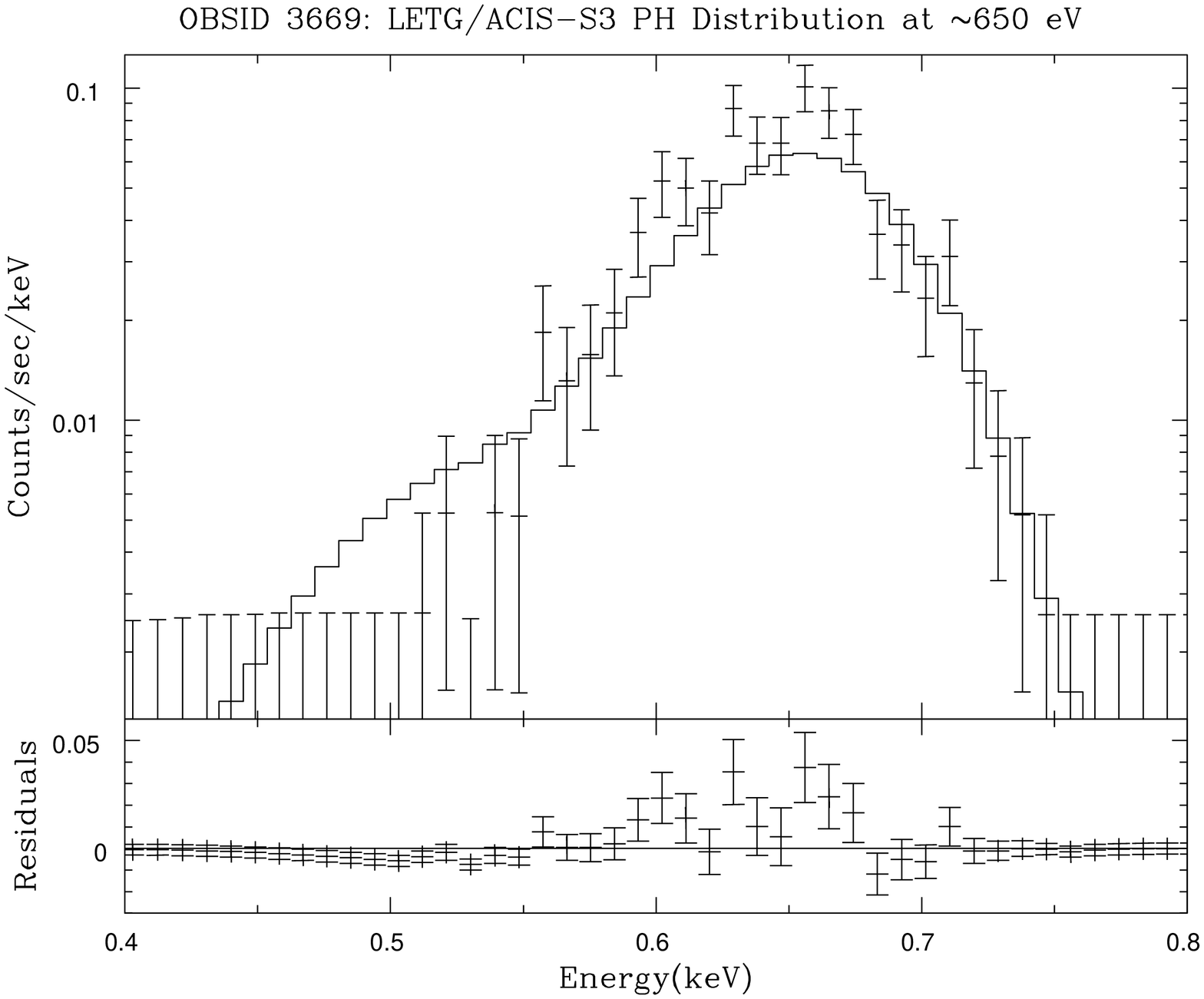}
   \includegraphics[width=2.7in,angle=270]{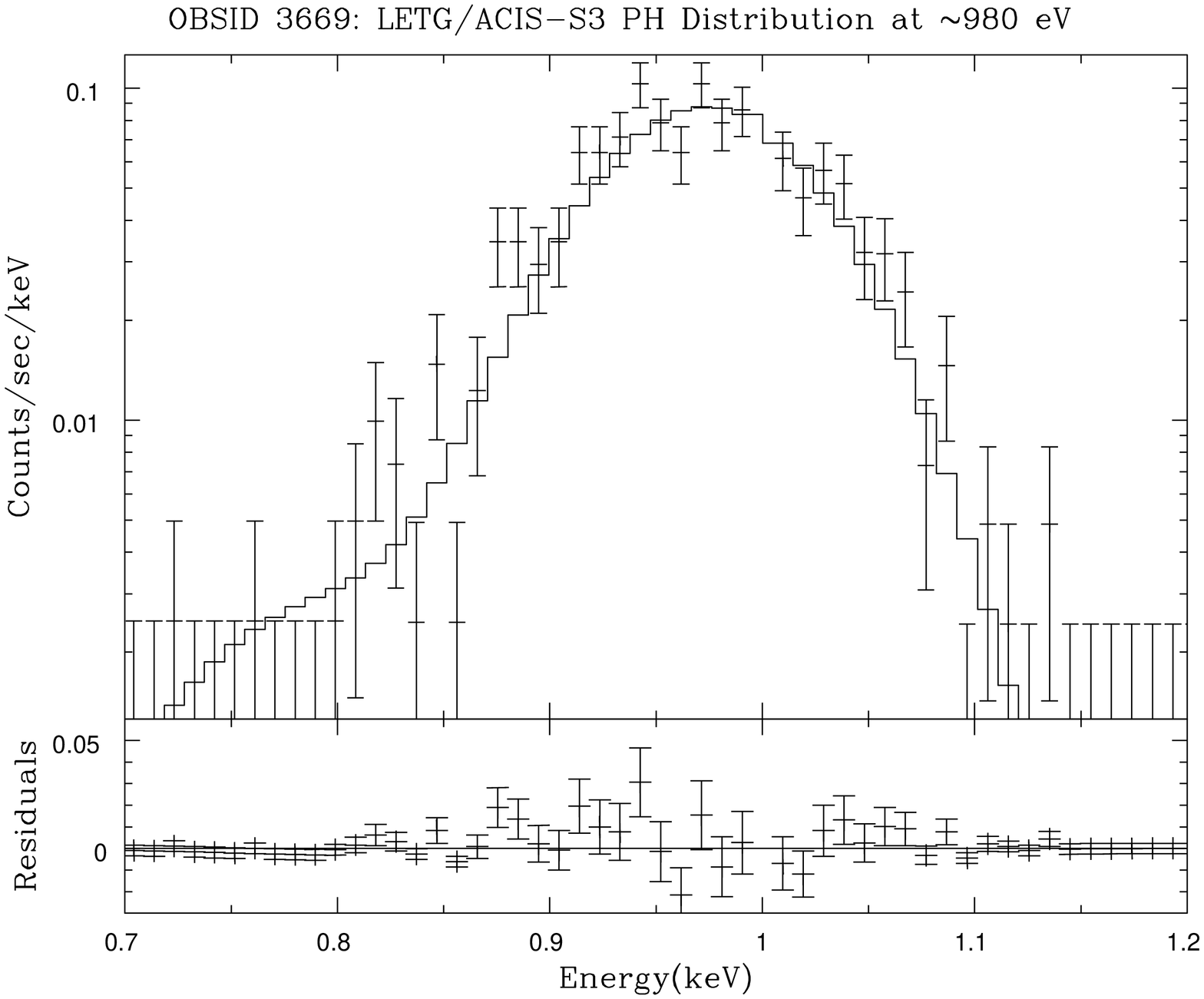}
   \end{tabular}
   \end{center}
   \vspace{-0.15in}
   \caption[nssredist] 
   { \label{fig:nssredist}
The PH distributions extracted from the region of the CCD
corresponding to energies from 0.636 to 0.670~keV (LEFT) and 0.943 to 
1.021~keV (RIGHT)
from the LETG/ACIS observation of PKS2155-304 (OBSID 3669).  The tail
of the spectral redistribution is not well-represented for energies
around 650~eV but is well represented for energies above 0.90~keV.
}
   \end{figure}

The PKS2155-304 LETG/ACIS data are also useful for investigating
the model of the spectral redistribution function.  Once again, we
can restrict our analysis to a very narrow range of energies by
selecting events by position on the CCD.  We have extracted data
from the location on the CCD which corresponds to an energy of
 $\sim0.65$~keV and $\sim0.97$~keV and fit these data with the existing 
response matrix for S3.  The data and model fits are displayed in 
Figure~\ref{fig:nssredist}.  The low-energy side of the 0.65~keV peak
is not well represented by the existing model.  It is clear that the
model overestimates the contribution of the tail.  The 0.97~keV data
are well-fitted by the model, both the peak and the low energy tail.
We will
be able to use the multiple calibration observations of PKS2155-304
to refine the spectral redistribution model of the S3 CCD at
low energies to be incorporated in an updated matrix of the CXO CALDB.
These data verify that the spectral redistribution function is
well-represented by the existing matrix for energies above 0.9 keV.

\section{APPLICATIONS TO CELESTIAL SOURCES}

We have been evaluating the calibration products using a variety
of celestial sources.  Since the contamination is on the filter and/or
CCDs, CXO observations with and without the gratings are affected.
  We have used 1E0102.2-7219 because it has a
soft, line-dominated spectrum which has been well-characterized by
the gratings on the CXO\cite{flanagan01} and 
{\em XMM-Newton}\cite{rasmussen00}. We have developed a
spectral model based on the lines observed in the gratings data
which contains 24 Gaussians for the lines, a two component absorption
(one for the Galactic component and one for the SMC), and a
bremsstrahlung for the continuum.  As mentioned in the earlier
sections, we have used the bright blazar PKS2155-304 which, even
though the intensity and spectral shape vary in time, can be
well-modeled by absorption plus a two component power-law. We have
also used the galactic SNR G21.5-0.9 which has a heavily-absorbed
power-law spectrum\cite{slane00}.  Finally, we have used the pulsar PSR 0656+14
which has a soft, continuum spectrum.

%
There are two SW tools which have been developed to model the
time-dependent effect of the absorption by the contamination layer.
One is an {\tt XSPEC} model, which would be used as an additional 
multiplicative component while fitting in {\tt XSPEC} (version 11.2 or above)
or {\tt SHERPA} and is called {\tt ACISABS}.
The other is a stand-alone program written in IDL and FORTRAN
called {\tt ACISABS.pro} and {\tt corr\_arf}, respectively, that modifies 
the effective area values in the arf file.
The {\tt ACISABS} tools calculate the transmission through 
the contamination layer assumed to consist of a hydrocarbon of
the form ${\rm C}_{n1}, {\rm H}_{n2}, {\rm O}_{n3}, {\rm N}_{n4}$, 
where, n1,n2,n3 and n4 
are input parameters and represent the number of atoms of 
the respective element in the hydrocarbon molecule.
The mass absorption coefficients of the contaminant
are calculated from the atomic scattering factor files 
provided at ``{\tt http://www-cxro.lbl.gov/optical\_constants/asf.html}''.
Both of these SW tools are available from the CXC contributed SW page:
``{\tt http://asc.harvard.edu/cont-soft/soft-exchange.html}''.
The results of this paper have made use of the corr\_arf program 
which assumes that C, H, N, and O are present in the contamination
layer in the ratio of 10:20:1:2, respectively.
The composition of the contaminant will probably need to be 
updated when the nature of the contaminant is better constrained 
with future observations.


The SW to correct events list for CTI is also available from the
CXC contributed SW page. It currently exists in the form of IDL code
which the user must install at their site.  The CXC has been
implementing the CTI correction in the standard processing of the
{\em Ciao} SW to be released in a future version.  For this paper, we
have used a development version of the {\em Ciao} SW which includes
the CTI correction algorithm.

\subsection{G21.5-0.9: HARD, CONTINUUM SOURCE}

As a first example, we present a source with a hard continuum spectrum
for which the fitted results do not vary significantly because of the
absorption by the contamination layer.
We have fit three observations of G21.5-0.9 separated by over a year
with and without the effective area correction.
All of these observations are on the S3 CCD near the nominal aimpoint.
We fit the data with an absorption plus power-law model.
Unfortunately, the ACIS configuration was different for the
observations. OBSID 1717 used a full-frame readout with a frametime
of 3.2~s and OBSIDs 1553 and 1554 used a frametime of 0.8~s with a
subarray readout.  Therefore, the pileup will be different in the 
observations.  The results and the 90\% confidence limits (CL) in 
Table~\ref{tab:g21.5tab} indicate that
there is a negligible difference between the fit with and without the
effective area correction for all three observations.  
 Therefore, any spectral results
derived from ACIS before July 2001 from a source with a hard,
continuum spectrum such as that of G21.5-0.9 will change by less
than 2\% when the effective area correction is applied.
The derived
flux values for the observations with the shorter frametime are
$\sim4\%$ higher, as one would expect if the pileup were reduced in
these observations. 


\begin{table}[h]
\centering

\caption[ ]{{\bf {Spectral Fit Results for G21.5-0.9 with and without
the effective area correction }}}
\label{tab:g21.5tab}
\begin{tabular}{|c|l|l|l|l|l|l|c|}
\hline
OBSID & DATE & E.A. & Frm & \NH\/($10^{22}~{\rm cm^{-2}}$) & Power-law 
& Norm~@~1~keV  & Flux ($\times10^{11}$)  \\
    & & Cor. & Tim  &  & Index & ($\times10^{-2}$) (${\rm photons}$  
& [0.5-9.0 keV]   \\
    & & & (s) & & & ${\rm /keV/cm^2/s}$)
& (${\rm ergs/cm^{2}/s}$)  \\
\hline
\hline
1717 &2000-05-23 &No & 3.2 & 2.25 [2.18,2.32] & 1.84 [1.78,1.89] & 2.02 [1.87,2.18] &  5.33 \\
1717 &2000-05-23 &Yes & 3.2 &2.22 [2.15,2.29] & 1.83 [1.78,1.89] & 2.01 [1.86,2.17]  & 5.34 \\
1553 &2001-03-18 &No & 0.8 &2.28 [2.22,2.34] & 1.86 [1.81,1.90] & 2.18
[2.04,2.33] &  5.55    \\
1553 &2001-03-18 &Yes & 0.8 &2.23 [2.17,2.29] & 1.85 [1.81,1.90] & 2.16 [2.02,2.31] &
5.58 \\ 
1554 &2001-07-21 &No & 0.8 &2.23 [2.18,2.30] & 1.82 [1.78,1.87] & 2.06
[1.92,2.21]   & 5.54 \\
1554 &2001-07-21 &Yes & 0.8 &2.18 [2.12,2.25] & 1.82 [1.77,1.87] & 2.04
[1.91,2.19]  &      5.57  \\
\hline

\end{tabular}

\end{table}

\subsection{1E0102.2-7219: SOFT, LINE-DOMINATED SOURCE}

\begin{figure}[h]
   \vspace{-0.30in}
   \begin{center}
   \begin{tabular}{c}
   \includegraphics[width=3.5in,angle=270]{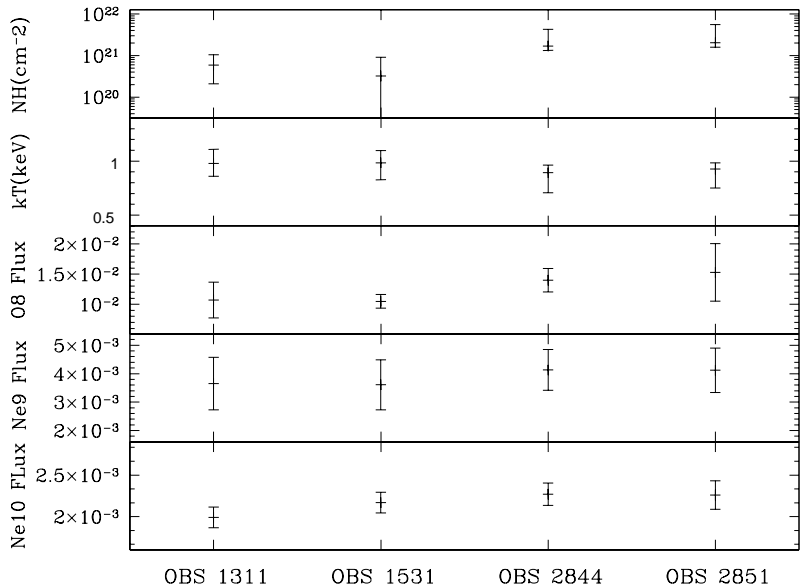}
   \end{tabular}
   \end{center}
   \vspace{-0.15in}
   \caption[e0102spres] 
   { \label{fig:e0102spres}
Spectral results from E0102 from the S3(BI) CCD for four observations 
spanning two years after applying the effective area correction. The
line fluxes are in units of ${\rm photons~cm^{-2}~s^{-1}}$.
}
   \end{figure}

For a source with a soft, line-dominated spectrum, the effective area
correction makes a large difference. We fit the four observations of
1E0102.2-7219 shown in Figure~\ref{fig:E0102_time} which show dramatic
differences at the lowest energies.  These observations used the
S3 detector and were within 0.5 arcminute of the on-axis aimpoint.
The spectrum is dominated by
strong lines of O and Ne.  We determined the flux in the 
O~{\small VIII}~Ly~$\alpha$, Ne~{\small IX}~triplet, and the
Ne~{\small X}~Ly~$\alpha$ lines in each of the four observations, 
in addition to the best-fit values for the \NH\/ and ${\rm kT}$.
The results with the 90\% CL are presented in 
Figure~\ref{fig:e0102spres}.  The \NH\/ values for OBSIDs 1311
and 1531 are in good agreement with estimates 
from optical measurements.  However, the \NH\/ values for OBSIDs
2844 and 2851 are significantly higher.  It is not clear if this is
because the absorption from the contamination layer has been
underestimated as a function of time or if a change in the detector 
response with time could
mimic an excess \NH\/.  The fitted values for the temperature and the
line fluxes are in agreement over this two year period.  There is a
hint that the increase in the derived flux for the 
O~{\small VIII}~Ly~$\alpha$ line might be correlated with the increase
in the derived value of the \NH\/, but the fluxes agree within the
90\% CL.  The derived fluxes for the Ne lines are in agreement with
each other for all four observations.

\subsection{PSR0656+14: SOFT, CONTINUUM SOURCE}

\begin{figure}[h]
   \vspace{-0.30in}
   \begin{center}
   \begin{tabular}{c}
   \includegraphics[width=3.5in,angle=270]{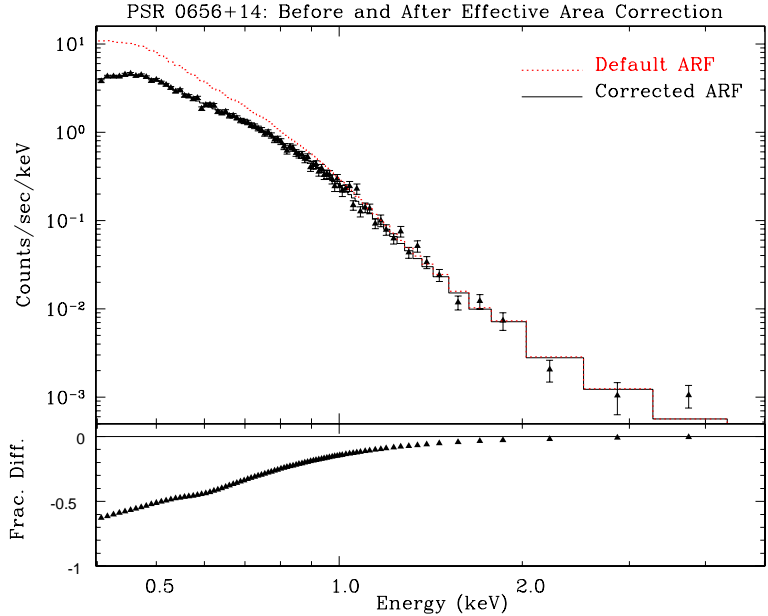}
   \end{tabular}
   \end{center}
   \vspace{-0.15in}
   \caption[psr0656] 
   { \label{fig:psr0656}
Spectrum of PSR0656+14 with the best-fit spectrum with the corrected
effective area plotted in black.  This model is then folded through 
the default
effective area curve (plotted in red) and the difference of the two
models is plotted in the lower panel.
}
   \end{figure}

We have also fit the spectrum of the nearby pulsar PSR~0656+14.
This source has a very soft spectrum which is only slightly modified
by the intervening interstellar medium.
ACIS was operated in continuous-clocking (CC) mode for this
observation.
Pavlov~\etal\/(2002)~\cite{pavlov02} have fit the spectrum with a
three component model consisting of two blackbody components and a
power-law component. Marshall~\& Schulz~(2002)\cite{marshall02}
analyzed the HRC/LETG spectrum of PSR~0656+14, which is unaffected
by the contamination layer and also provides the highest resolution
spectral data acquired for this object, and derived parameters for the
two blackbody components which are consistent with the ACIS CC mode
results.  
We fit the spectrum with and without the effective area correction.  
Figure~\ref{fig:psr0656} displays the best fit with the effective area
correction and the same model folded through the default effective
area curve.
%
The temperatures of the blackbody components changed little between the fits 
but the normalizations changed significantly, leading to a large
difference in the derived flux.  
The fitted values of the flux of these two fits are shown in 
Table~\ref{tab:psr0656}.  We have restricted the fit to
energies from 0.4~keV to 5.0~keV due to the uncertainties in the
low-energy response of ACIS below 0.4~keV but we have computed the
flux from 0.24~keV to 3.0~keV to compare to previous results.
These results demonstrate that the derived flux can be different by 
as much as a factor of three for a soft, continuum spectrum.


\begin{table}[h]
\centering

\caption[ ]{{\bf {Spectral Fit Results for PSR~0656+14 with and without
the effective area correction }}}
\label{tab:psr0656}
\begin{tabular}{|c|l|}
\hline
Effective Area Correction & Flux [0.24-3.0 keV]  (${\rm ergs/cm^{2}/s}$) \\
\hline
\hline
No & $0.70\times10^{11}$ \\
Yes & $2.60\times10^{11}$ \\
\hline

\end{tabular}

\end{table}

\subsection{1E0102.2-7219: CTI CORRECTION}

\begin{figure}[h]
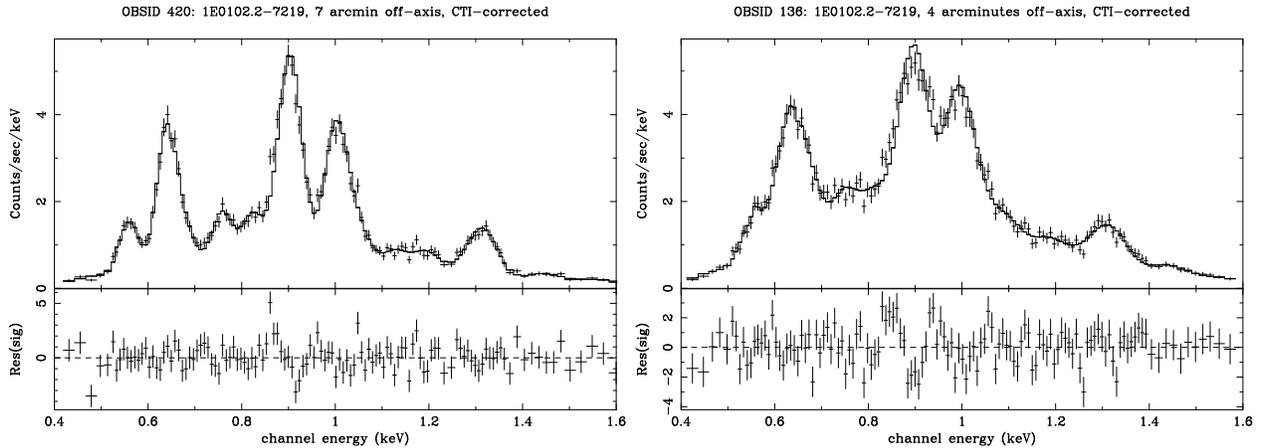

   \begin{center}
   \begin{tabular}{c}
   \includegraphics[width=2.3in,angle=270]{plucinsky_spie02_fig10a.ps}
   \includegraphics[width=2.3in,angle=270]{plucinsky_spie02_fig10b.ps}
   \end{tabular}
   \end{center}
   \vspace{-0.15in}
   \caption[e0102fi1] 
   { \label{fig:e0102fi1}
Spectra of 1E0102.2-7219 extracted from the I3 CCD from near
the framestore (LEFT) OBSID 420 and in the middle of the CCD
(RIGHT) OBSID 136.  The CTI correction and the effective area
correction have been applied.
}
   \end{figure}

\begin{figure}[h]
   \begin{center}
   \begin{tabular}{c}
   \includegraphics[width=2.4in,angle=270]{plucinsky_spie02_fig11a.ps}
   \includegraphics[width=2.4in,angle=270]{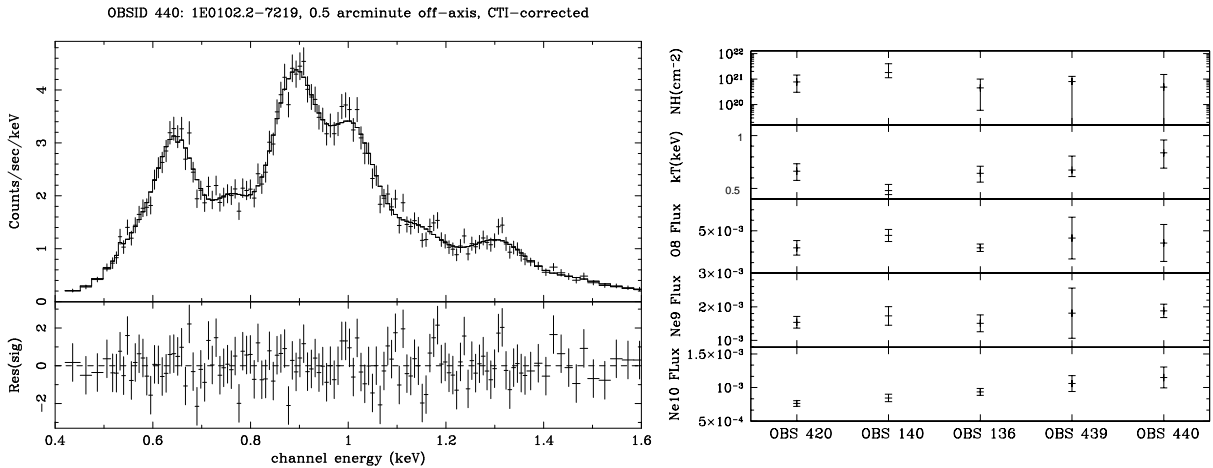}
   \end{tabular}
   \end{center}
   \vspace{-0.15in}
   \caption[e0102fi2] 
   { \label{fig:e0102fi2}
(LEFT) Spectrum of 1E0102.2-7219 extracted from the I3 CCD from near
the aimpoint OBSID 440. (RIGHT) The derived values from the fits of
the five observations which span the I3 CCD.  There is excellent
agreement among the observations after the CTI correction has
been applied and the effective area correction.
}
   \end{figure}

Five observations of 1E0102.2-7219 were executed at different
{\tt chipy} locations on the I3 CCD from near the nominal 
aimpoint to the framestore region. The spacing between
observations was $\sim150$ rows.  We have
analyzed these observations to investigate the performance of the CTI
correction algorithm at low energies as the source is moved across
the CCD.  We fit these data with the same model described earlier.
We applied both the CTI correction and the effective area correction 
to the I3 observations.  Three of the five spectra are plotted in
Figures~\ref{fig:e0102fi1} and~\ref{fig:e0102fi2}.  The three
observations which
are plotted are OBSIDs 420, 136, and 440, which are closest to the
framestore, in the middle of the CCD, and closest to the aimpoint
respectively.
The spectrum for OBSID 420 exhibits spectral resolution at nearly
the pre-launch, undamaged value. Notice how the O~{\small VII} triplet
is resolved from the O~{\small VIII}~Ly~$\alpha$ line and the
Ne~{\small IX} triplet is resolved from the Ne~{\small X}~Ly~$\alpha$
line.  OBSID 136 demonstrates the spectral resolution available in
the middle of the CCD.  The lines have clearly started to blend
together. Finally, OBSID 440 demonstrates the resolution available
at the top of the CCD.  Similar to the exercise for 1E0102.2-7219
on S3, we have determined the fitted parameters of the \NH\/,
kT, O~{\small VIII}~Ly~$\alpha$ line flux, the Ne~{\small IX} triplet
flux and the Ne~{\small X}~Ly~$\alpha$ line flux for these five
observations.  The values are
plotted versus OBSID in Figure~\ref{fig:e0102fi2}.  There is
excellent agreement among the observations for all the fitted
parameters except for the \NH\/ and kT for OBSID 140.  These
results demonstrate that after applying the CTI correction and
the effective area correction, consistent results can be 
achieved for a source with a soft, line-dominated spectrum as a
function of time on the I3 CCD.

\subsection{CONCLUSIONS}

   The flight spectral response of the ACIS instrument is
well-characterized in the 1.5~keV to 8.0~keV bandpass.  The current
calibration products provide for consistent results between
observations of the same source at different epochs and at
different detector positions within the measurement uncertainties.  
The uncertainties in the calibration
below 1.5~keV are significantly larger and are dominated by the
temporal variation of the spectral resolution and the detection
efficiency.  The level of uncertainty depends strongly upon the
spectral shape of the source and the precise energy of interest.
We have demonstrated that the application of a time-dependent
effective area correction and a CTI correction improve significantly
the consistency of the results below 1.5~keV.

\acknowledgments     

This work was supported by NASA contract NAS8-39703.\\
We thank many of our colleagues on the CXO project who have
contributed directly or indirectly to this work.  We thank
Mark Bautz, Bev LaMarr, Peter Ford,  
Dan Schwartz, Leisa Townsley,
George Pavlov, Konstantin Getman, Martin Weisskopf, Steve
O'Dell, Allyn Tennant, Ron Elsner and all members of the ACIS
instrument team and CXO Project Science who have contributed to this
effort.  We thank Slava Zavlin for granting us permission to use the
observation of PSR0656+14 which is still proprietary at this time.
For anyone we have forgotten,
we apologize.


\bibliography{plucinsky_spie02_ver2}   

\begin{thebibliography}{10}

\bibitem{weiss00}
M.~C. Weisskopf, H.~D. Tananbaum, L.~P. Van~Speybroeck, and S.~L. O'Dell,
  ``Chandra x-ray observatory (cxo): overview,'' in {\em X-Ray Optics,
  Instruments, and Missions III},  J.~E. Truemper and B.~Aschenbach, eds., {\em
  Proc. SPIE} {\bf 4012}, p.~2, 2000.

\bibitem{weiss02}
M.~C. Weisskopf, B.~Brinkman, C.~Canizares, G.~Garmire, S.~Murray, and L.~P.
  Van~Speybroeck, ``An overview of the performance and scientific results from
  the chandra x-ray observatory,'' {\em Publications of the Astronomical
  Society of the Pacific} {\bf 114}, p.~1, jan 2002.

\bibitem{bautz98}
M.~Bautz, M.~Pivovaroff, F.~Baganoff, T.~Isobe, S.~Jones, S.~Kissel, B.~Lamarr,
  H.~Manning, G.~Prigozhin, G.~Ricker, J.~Nousek, C.~Grant, K.~Nishikida,
  F.~Scholze, R.~Thornagel, and G.~Ulm, ``X-ray ccd calibration for the axaf
  ccd imaging spectrometer,'' in {\em X-{R}ay {O}ptics, {I}nstruments, and
  {M}issions},  R.~B. Hoover and A.~B.~W. II, eds., {\em Proc. SPIE} {\bf
  3444}, p.~210, 1998.

\bibitem{garmire92}
G.~Garmire, G.~Ricker, M.~Bautz, B.~Burke, D.~Burrows, S.~Collins, J.~Doty,
  K.~Gendreau, D.~Lumb, and J.~Nousek, ``The axaf ccd imaging spectrometer,''
  in {\em American Institute of Aeronautics and Astronautics Conference},
  p.~8, 1992.

\bibitem{prig00a}
G.~Prigozhin, S.~Kissel, M.~Bautz, C.~Grant, B.~LaMarr, R.~Foster, G.~Ricker,
  and G.~Garmire, ``Radiation damage in the chandra x-ray ccds,'' in {\em
  X-{R}ay {O}ptics, {I}nstruments, and {M}issions},  J.~E. Truemper and
  B.~Aschenbach, eds., {\em Proc. SPIE} {\bf 4012}, p.~720, 2000.

\bibitem{prig00b}
G.~Prigozhin, S.~Kissel, M.~Bautz, C.~Grant, B.~LaMarr, R.~Foster, and
  G.~Ricker, ``Characterization of the radiation damage in the chandra x-ray
  ccds,'' in {\em X-{R}ay {O}ptics, {I}nstruments, and {M}issions},  K.~A.
  Flanagan and O.~H. Siegmund, eds., {\em Proc. SPIE} {\bf 4140}, p.~123, 2000.

\bibitem{town00}
L.~K. Townsley, P.~S. Broos, G.~P. Garmire, and J.~A. Nousek, ``Mitigating
  charge transfer inefficiency in the chandra x-ray observatory advanced ccd
  imaging spectrometer,'' {\em Astrophysical Journal} {\bf 534},
  pp.~L139--LL142, 2000.

\bibitem{town02a}
L.~K. Townsley, P.~S. Broos, J.~A. Nousek, and G.~P. Garmire, ``Modeling charge
  transfer inefficiency in the chandra advanced ccd imaging spectrometer,''
  {\em Nuclear Instruments and Methods} {\bf 486}, p.~751, 2002.

\bibitem{odell02}
S.~O'Dell and A.~Tennant {\em private communication} , 2002.

\bibitem{wilms00}
J.~Wilms, A.~Allen, and R.~McCray, ``On the absorption of x-rays in the
  interstellar medium,'' {\em Astrophysical Journal} {\bf 542}, p.~914, 2000.

\bibitem{ppp01}
P.~P. Plucinsky, L.~Townsley, P.~S. Broos, R.~J. Edgar, and S.~N. Virani, ``The
  low-energy spectral response of the acis ccds on the chandra x-ray
  observatory,'' in {\em High Energy Universe at Sharp Focus: Chandra Science},
   E.~M. Schlegel and S.~D. Vrtilek, eds., p.~391, 2001.

\bibitem{bautz99}
M.~Bautz, G.~Prigozhin, M.~Pivovaroff, S.~Jones, S.~Kissel, and G.~Ricker,
  ``X-ray ccd response functions, front to back,'' {\em Nuclear Instruments and
  Methods A} {\bf 436}, p.~40, 1999.

\bibitem{town02b}
L.~K. Townsley, P.~S. Broos, G.~Chartas, E.~Moskalenko, J.~A. Nousek, and G.~G.
  Pavlov, ``Simulating ccds for the chandra advanced ccd imaging
  spectrometer,'' {\em Nuclear Instruments and Methods} {\bf 486}, p.~716,
  2002.

\bibitem{flanagan01}
K.~A. Flanagan, C.~R. Canizares, D.~S. Davis, D.~Dewey, J.~C. Houck, and M.~L.
  Schattenburg, ``Ionization structure and the reverse shock in eo102-72,'' in
  {\em AIP Conf. Proc. 565: Young Supernova Remnants},  p.~226, 2001.

\bibitem{rasmussen00}
A.~P. Rasmussen, E.~Behar, S.~M. Kahn, J.~W. den Herder, and K.~van~der Heyden,
  ``The x-ray spectrum of the supernova remnant 1e 0102.2-7219,'' {\em
  Astronomy \& Astrophysics} {\bf 365}, p.~L231, 2001.

\bibitem{slane00}
P.~Slane, Y.~Chen, N.~S. Schulz, F.~D. Seward, J.~P. Hughes, and B.~M.
  Gaensler, ``Chandra observations of the crab-like supernova remnant
  g21.5-0.9,'' {\em Astrophysical Journal} {\bf 533}, p.~L29, 2000.

\bibitem{pavlov02}
G.~Pavlov, V.~Zavlin, and D.~Sanwal, ``Thermal radiation from neutron stars,''
  in {\em The 270-th WE-Heraeus Seminar on Neutron Stars, Pulsars and Supernova
  Remnants},  H.~L. W.~Becker and J.~Truemper, eds., {\em MPE Report} {\bf
  278}, p.~273, 2002 (astro-ph/0206024).

\bibitem{marshall02}
H.~L. Marshall and N.~S. Schulz, ``Using the high-resolution x-ray spectrum of
  psr b0656+14 to constrain the chemical composition of the neutron star
  atmosphere,'' {\em Astrophysical Journal} {\bf 574}, pp.~377--381, 2002.

\end{thebibliography}
\bibliographystyle{spiebib}   

\end{document}